\documentclass[preprint,aps,prb]{revtex4}
\usepackage{graphicx}

\renewcommand{\appendix}{
 \setcounter{section}{0}
 \setcounter{equation}{0}%
 \renewcommand{\thesection}{{APPENDIX} \Alph{section}}
 \renewcommand{\theequation}{\Alph{section}.\arabic{equation}}%
}

\begin{document}
\preprint{LA-UR-03-6169}
\title{\bf Theory of polyelectrolytes in solvents}
\author{Shirish M. Chitanvis}
\affiliation{
Theoretical Division, 
Los Alamos National Laboratory\\
Los Alamos, New Mexico\ \ 87545\\}

\date{\today}

\begin{abstract}
Using a continuum description,
we account for fluctuations in the ionic solvent surrounding a
Gaussian, charged chain and derive an effective short-ranged potential between the
charges on the chain.
This potential is repulsive at short separations and attractive at
longer distances.
The chemical potential can be derived from this potential.
When the chemical potential is positive, it leads to a melt-like
state.
For a vanishingly 
low concentration of segments, this state exhibits scaling behavior
for long chains.
The Flory exponent characterizing the radius of gyration for long
chains is calculated to be approximately $0.63$, close to the
classical value obtained for second order phase transitions.
For short chains, the radius of gyration varies linearly with $N$, the
chain length, and is sensitive to the parameters in the interaction potential.
The linear dependence on the chain length $N$ indicates a
stiff behavior.
The chemical potential associated with this interaction changes sign,
when the screening length in the ionic solvent exceeds a critical value.
This leads to condensation when the chemical
potential is negative.
In this state, it is shown using the mean-field
approximation that spherical and toroidal condensed shapes can be
obtained.
The thickness of the toroidal polyelectrolyte is studied as a function
of the parameters of the model, such as the ionic screening length.
The predictions of this theory should be amenable to experimental verification.
\end{abstract}
\maketitle


\section {Introduction}

A distinguishing feature of biopolymers is the presence of charges
along the chain and
their subsequent interaction with each other and ionic aqueous solvents.  Yet another
aspect is the semi-rigidity of some biopolymers.  In this sense,
biopolymers differentiate themselves from polymer melts, which are
neutral, and most theoretical treatments involve the treatment of
flexible polymers. Furthermore, these theoretical treatments of melts
almost exclusively use models of a self-excluded volume nature.
A general treatment of electrostatic interactions between charges on a
polymer is crucial
in order to extend the validity of theoretical models to bio-polymers,
polymers exhibiting counter-ion condensation in ionic solvents, etc.
\cite{frey_1,frey_2,frey_3,hinner}.

This paper lays a field-theoretic foundation for a theoretical
treatment of charged polymers in ionic solutions. 
Technically, the theory applies to flexible polyelectrolytes.
However, we have argue that by choosing the Kuhn length
to be the correlation length of a semi-rigid polymer, it should be
possible to address the physics of semi-flexible polymers.
The intent is not only to understand the
behavior of the radius of gyration with chain length, but also to
gain a basic knowledge of condensation phenomena.
Our calculations of the radius of gyration generalize standard
treatments for melts. A   $~$mean-field approach was employed to develop a
phase diagram for condensation/non-condensation of charged
polymers. In this endeavor, we found the recent experimental paper by Butler et
al\cite{butler} very useful.

For the case of excluded volume interactions in polymer melts, it was shown by
deGennes\cite{deG1} and others\cite{cloizeaux,oono,shirish0} that in the limit
of low number concentration of 
monomers, the physics of polymers 
is analogous to the onset of a second order phase transition.
It will be shown that a similar situation ensues when considering
polyelectrolytes.
The Flory exponent in this case, for long chains, is argued to be
identical to the classical case.
This is in distinction to the arguments\cite{barrat} given by
Katchalsky\cite{kat} and Flory\cite{flory-el}
 for polyelectrolytes, where a long-range,
unscreened Coulomb interaction was employed and the effects of any
solvent were ignored. 

This paper develops a model which begins with a screened Coulomb
interaction between native electron charges and adsorbed counter-ions
along the segments of a Gaussian chain.
This screening length is independent of the solvent properties.
These charges also interact with the surrounding ionic solvent,
treated as a continuum.
The theory in this paper continues the development of a
functional integral technique\cite{shirish0} which begins with the
formulation of a system 
of many homopolymeric flexible chains, and whose segments interact
through a finite-range potential.

The functional integral technique utilized in this paper is different
from the coarse-graining method employed by Fredrickson et
al.\cite{fred}
Fredrickson et al use the number density of the polymer segments as the
order parameter.
The advantage afforded by this technique is that the
interaction term becomes quadratic and hence trivial to calculate.
However the entropy, or free-chain part of the energy functional
becomes non-linear, complex and hence presents a computational challenge.
The order parameter introduced in our paper is a probability amplitude whose
absolute square corresponds to the number density of segments.
The free-chain portion of the energy functional is linear in this
approach, while the interaction term remains understandably non-linear
and difficult computationally.
The positive semi-definite form of our energy functional may also offer
some computational advantages.
Our formalism can be viewed as a generalization of the
self-consistent field theory used by Curro's group.\cite{deG1,curro}
The model of interaction used in this paper differs somewhat
from that utilized by Curro et al.
Our approach is most similar to that of Kleinert\cite{kleinert,ferrari}, but the scope of our
study is broader.

Functional integration is used to integrate over the fluctuations in
the ionic solvent, and derive an effective interaction between the
charges on the chain.
This interaction is repulsive at short distances, and attractive for
larger separations.
The chemical potential $\mu$ corresponding to this interaction can be
computed. 
The chemical potential changes sign, depending on the parameter regime.
Using the interpretation of $\mu$ as the energy required to add a
segment (monomer) to the system, one associates 
$\mu > 0$ with melt-like behavior, and
$\mu < 0$ with a condensed state. 
Indeed it will be shown that when $\mu < 0$, it is
possible for a single strand of polyelectrolyte to condense into a
toroidal shape. Spherical shapes are also possible.

\section{The basic formalism}

We reproduce here briefly for completeness, the functional integral
formulation of a homopolymer chain.
For the record, it must be stated  that this formalism differs from
the standard mapping of the self-avoiding walk of a single chain onto
a $\phi^4$ field theory.\cite{deG1,cloizeaux}
It can be generalized in a straightforward way to deal with many
chains. Technically speaking, the model is restricted to flexible
chains.
However, if one treats the segment length as a Kuhn length, i.e. as
an effective distance which represents the correlation length of a
semi-flexible polymer, then one expects this model to reproduce the
overall features of a semi-rigid chain as well.

The probability distribution $G_0(1,2;n)$ for a single
Gaussian chain may be represented 
by\cite{doi-ed}: 

\begin{eqnarray}
G_0(1,2;n)  &&= \left({3\over 2 \pi b^2 |n|}\right)^{3/2} \exp\left(-{3\
      |\vec R_1 - \vec R_2|^2 \over 2 |n|b^2}\right)  \nonumber\\
&& \sim \int_{\vec R_2} ^{\vec R_1} {\cal D}\vec R(n')
\exp-\left[\left({3\over 2 b^2}\right)  \int_0 ^{n} dn' \left({\partial
      \vec R(n') \over \partial n' }^2 \right)\right]\nonumber\\
&&\equiv \langle 1,n\vert \left[ \partial_{n'} - \left({b^2
      \over 6}\right) 
  \nabla^2 \right]^{-1} \vert 2,0 \rangle
\label{one}
\end{eqnarray}

where $b$ is the Kuhn length of the polymer as discussed above, and where $\partial_{n'}
\equiv {\partial \over \partial n}$.
This expression is obtained by considering only the entropy of a
flexible chain.

$G_0(1,2;n)$ can also be thought of as the Green's function for the
diffusion operator in three dimensions, as indicated by the last of
Eqns. \ref{one}.

Alternatively, one knows from methods in functional integration
that\cite{kaku}:

\begin{eqnarray}
&& \langle 1,n\vert \left[ \partial_{n} - \left({b^2\over6}\right)  
  \nabla^2 \right]^{-1} \vert 2,0 \rangle ~\sim~ \int {\cal D}^2\psi
~\psi^*(\vec R_1,n) 
\psi(\vec R_2,0)  \exp-[\beta {\cal F}_0] \nonumber\\  
&&\beta {\cal F}_0 = \int dn' 
d^3x~\psi^*(\vec x,n') \left[{\partial_{n'}} 
  -\left({b^2\over6}\right) \nabla^2 \right] \psi(\vec x,n')
\label{map}
\end{eqnarray}
where ${\cal D}^2 \psi \equiv {\cal D} \psi^* {\cal D} \psi$,
$ \beta = {1 \over k_B T} $,
$k_B$ is Boltzmann's constant and $T$ is the temperature.
Thus we have another way of thinking about a system of
flexible polymers, in terms of a probability amplitude $\psi(\vec x,n)$ and an energy
functional $\beta {\cal F}_0$ which is isomorphic to one that describes
diffusion.  
Here $(\vec x,n)$ labels the location $\vec x$ in physical
space, of the $n$-th segment of a chain.
A similar approach has been advocated by Kleinert.\cite{kleinert,ferrari}
Following the convention in quantum field theory, $\vert
\psi(\vec x,n)\vert^2$ is interpreted as the probability of finding a
polymer  
segment at a given location in space.

The main advantage of the functional path integral formalism is that
one can  model more easily interactions in systems with large numbers
of polymers by adding an interaction term as shown below:

\begin{eqnarray}
\beta {\cal F}_0 &&\to \beta {\cal F}=
\beta {\cal F}_0 + \beta \Delta {\cal F}
\nonumber\\
\beta \Delta {\cal F} &&= \int d^3x d^3x'~dn dn' \left(
  V(\psi(\vec x,n) ,\psi^*(\vec x,n')\right)
\label{interact0}
\end{eqnarray}
where $V(\psi(\vec x,n),\psi^*(\vec x', n'))$ represents the interaction between
segments.  Note that in general, on physical grounds, $n \ne n'$,
as different configurations of the strand permit segments from
different parts of the chain to be in the proximity of each other.

The form chosen for ${\cal V}$ in this paper is:

\begin{equation}
{\cal V}(\psi(\vec x,n) ,\psi^*(\vec x,n')) = \left({1 \over 2}\right)~
   \vert \psi(\vec x,n) \vert^2 ~U_{n,n'}(\vert \vec x-\vec x'
   \vert)~\vert \psi(\vec x',n') \vert^2 
\label{morse}
\end{equation}

Here $U_{n,n'}(\vec x - \vec x')$ represents a repulsive-attractive
short-ranged interaction between monomers.
The precise form of the potential useful for describing polyelectrolytes will be
discussed in detail the next section.

Upon extremizing the function $\beta {\cal F}$ with respect to
$\psi^*$, one obtains a non-linear diffusion equation: 

\begin{equation}
{\partial \psi(\vec x,n) \over \partial n} -  \left({b^2\over 6}\right)\nabla^2 \psi(\vec x,n)+
\int_0^N dn'~\int d^3x'  |\psi(\vec x',n')|^2 U_{n,n'}(|\vec x-\vec x'|)~\psi(\vec x,n) = 0
\label{hf}
\end{equation}

This equation is also analogous to the single-electron Hartree-Fock
equation in quantum mechanics, and applies to a {\em single} polymer strand.

If a generalization to many strands (several electron orbitals in
quantum mechanics) is required, then the requisite form would be:

\begin{equation}
{\partial \Psi_i(\vec x,n) \over \partial n} - \left({b^2\over 6}\right)\nabla^2 \Psi_i(\vec x,n)+
\sum_{j=1}^{N_c}~\int_0^N dn'~\int d^3x'  |\Psi_j(\vec x',n')|^2
U_{n,n'}(|\vec x-\vec x'|)~\Psi_i(\vec x,n) = 0
\label{hfm}
\end{equation}

where the subscript on the amplitude indicates a chain label, $N_c$
being the total number of chains in the system.
The rest of the paper will focus primarily on the physics of a single chain.

We shall now show how a connection can be made to standard results in
polymer theory.
The correlation function ${\cal G}_0(k,p)$, $p$ being the separation
between any two segments on the Gaussian chain can be obtained by using the Fourier
transform of the Gaussian distribution function discussed earlier in
the section.  This Fourier transform shall be denoted by 
$\hat G_0(k,\omega)$:

\begin{eqnarray}
\hat G_0(k,\omega) &&={1 \over -i \omega + (b k)^2/6} \nonumber\\ 
{\cal G}_0(k,p) &&= \int_{-\infty}^{\infty} {d \omega \over 2 \pi}\
\exp(-i \omega p) ~\hat G_0(k,\omega) 
\label{corr1a}
\end{eqnarray}

A simple contour integration, performed by closing the contour in the
complex $\omega$ plane in the clockwise direction leads to:

\begin{equation}
{\cal G}_0(k,p)=\exp(-\vert p\vert~(b k)^2/6)
\label{corr1}
\end{equation}

The radius of gyration $R_g^0(p)$ can be defined as:

\begin{eqnarray}
R_g^0(p) &&= \left({\partial \vert\ln({\cal G}_0(k,p))\vert^{\nu_0}\over \partial \
k}\right)_{k=0}\nonumber\\
&&= (b/\sqrt 6) \sqrt p \nonumber\\ {1 \over \nu_0} &&= {\rm power~of~k~}\equiv 2
\label{rg00}
\end{eqnarray}

The experimentally accessible structure factor associated with the
Gaussian chain of $N$ segments can defined in the following fashion:

\begin{eqnarray}
&&\hat S_0(k) = N^{-1}~\int_0^N~\int_0^N dm~dn~{\cal G}_0(k,\vert
m-n\vert) \nonumber\\
&&\approx N~(1- (1/3)~ k^2 R^0_g(N)^2) + {\cal O}(k^4)
\end{eqnarray}

$R_g^0(p)$ can also be defined in terms of the structure factor:

\begin{eqnarray}
R_g^0(N) &&=\left({\partial \left[3~\vert\ln(\hat
      S_0(k))\vert\right]^{\nu_0}\over \partial k}\right)_{k=0}\nonumber\\ 
&&= \left({b \over \sqrt 6}\right)~\sqrt N \nonumber\\
{1 \over \nu_0} &&= {\rm power~of~k~}\equiv 2
\label{rg0}
\end{eqnarray}



In the presence of interactions, $\hat G_0(k,\omega)$ gets
renormalized in the usual way, and an associated {\em dressed}
correlation function ${\cal G}(k,N)$ and dressed structure
factor $\hat S(k)$ can be defined analogously:

\begin{eqnarray} 
\hat G_0(k,\omega) &&\to (\hat G_0^{-1}(k,\omega) - \Sigma(k,\omega))^{-1}\nonumber\\
{\cal G}(k,N) &&= \int {d \omega \over 2 \pi} \exp(-i \omega N) ~
\hat G(k,\omega) \nonumber\\
\hat S(k) &&= N^{-1}~{\cal G}(0,N)^{-1}~\int_0^N~\int_0^N dm~dn~{\cal G}(k,\vert
m-n\vert)
\label{inter1}
\end{eqnarray}
where a normalization factor has been included in the definition of
the structure factor.

The radius of gyration  $R_g(N)$ in the presence of interactions can
be defined following the ideas for the free chain in terms of the
dressed correlation function:

\begin{eqnarray}
R_g(N) &&= \left({\partial \vert\ln({\cal G}(k,N){\cal G}(0,N)^{-1})\vert^{\nu}\over \partial
k}\right)_{k=0}\nonumber\\
{1 \over \nu} &&= {\rm power~of~k~}
\label{rg1}
\end{eqnarray}
where a normalizing factor has been inserted on the right hand side of
Eqn.\ref{rg1} for completeness.
Equivalently:

\begin{eqnarray}
R_g(N) &&= \left({\partial \left[3~\vert \ln(\hat S(k) \hat S(0)^{-1})\vert\right]^{\nu}\over
    \partial k}\right)_{k=0} \nonumber\\
{1 \over \nu} &&= {\rm power~of~k~in~the~argument~of~}{\hat S}(k,N)
\label{rg1a}
\end{eqnarray}


\section{Polyelectrolyte in a solvent}

Let us suppose that the net charge density on any segment along the chain is
given by sum of negative and positive charge densities on the chain: 
$\rho_t(\vec x,n)=\rho_-(\vec x,n)+\rho_+(\vec x,n)$, with $\rho_- <
0$, $\rho_+ > 0$.
Then the total free energy which describes the interaction between
these charges and the electrolytic solution is given by three terms,
describing the interaction of charges along the chain, the solvent,
and the interaction between the polymer and the solvent, respectively:

\begin{eqnarray}
&&\beta {\cal F}_0 \to \beta {\cal F} \nonumber\\
&&= \beta \left[{\cal F}_0 +\
   {\Delta \cal F}_{polymer} + {\Delta \cal F}_{solvent} + {\Delta \cal F}_{polymer-solvent}\right]
\nonumber\\ 
&&\beta {\Delta \cal F}_{polymer} = \left({\beta\over
      2}\right) \int d^3x~d^3x'~dn~dn'\
\rho_t(\vec x,n)~{\exp\left(-\kappa_e\vert \vec x - \vec x' \vert\right) \over\vert
  \vec x - \vec x' \vert}~\rho_t(\vec x',n') \nonumber\\ 
&&\beta {\Delta \cal F}_{solvent} = \left(\beta \epsilon \over 8 \pi \right) \int d^3x
\left[ \vert \vec \nabla \phi(\vec x) \vert^2 + \kappa^2 \phi(\vec
  x)^2 \right] \nonumber\\ 
&&\beta {\Delta \cal F}_{polymer-solvent} = - \beta \int d^3x~dn \rho_t(\vec x,n)~\phi(\vec x)
\label{interact}
\end{eqnarray}
where $\phi(x)$ is the electrostatic potential of the ionic solvent,
$\kappa_e=1/\lambda_e$ is the inverse screening length of the polymer,
$\kappa=1/\lambda$ is the inverse screening length in the solvent,
where $\epsilon$ is the dielectric constant of the solvent
($\epsilon\approx 80$ for water).
The reason for assuming the screening lengths to be distinct for the
chain ($\lambda_e$) and the solvent ($\lambda$) is that even in the absence of an ionic solvent,
when the adsorbed charge is null, the conformations of the negatively
charged Gaussian chain can screen the Coulomb interaction between the native
charges on the chain.

The charge distribution on the chain is given by:

\begin{eqnarray}
\rho_-(\vec x,n) &&= q_-(n)~\vert \psi(\vec x,n)\vert^2,~q_-(n) < 0
\nonumber\\
\rho_+(\vec x,n) &&= q_+(n)~\vert \psi(\vec x,n)\vert^2,~q_+(n) > 0
\label{chg}
\end{eqnarray}

Here $q_-(n)$ refers to the native negative charge on the backbone of
the polymer, and $q_+(n)$ refers to the positive counter-ions that may
be adsorbed onto the chain.
One can conclude from this that if the negative native charges are singly
charged, then the adsorbed counter-ions would have to be multiply
charged, as otherwise the net charge would be zero, and we would be
reduced to studying a neutral melt. It also allows for the
possibility that the counter-ions are docked in a physically different
location along the chain than the native charges, which would allow
for local regions of attraction to be created.

Beyond these inferences, one notes that if the above function
Eqn.\ref{interact} were to be extremized with respect to the the scalar
potential $\phi$, one obtains the standard Debye-Huckel
equation, valid in the limit of low ionic strengths. 
Thus our paper goes beyond the standard, mean-field Debye-Huckel
model by considering fluctuations around that
approximation in the full energy functional given by Eqn. \ref{interact}. 
These fluctuations will be shown shortly to lead to an attractive
component of the correlations between like charges along the chain.

Note that the electrostatic potential appears quadratically in
the functional.
One can now integrate over solvent degrees of freedom, in particular,
accounting for deviations from the Debye-Huckel mean-field limit to obtain an
effective functional:

\begin{eqnarray}
&&Z = \int {\cal D}^2 \psi~{\cal D}\phi \exp\left[-\beta {\cal F}\right] \nonumber\\ 
&& \sim \int d^2 \psi \exp\left[-\beta {\cal F}_{effective}\right] \nonumber\\
&&\beta {\cal F}_{effective} = \beta \left[ {\cal F}_0 + {\cal F}' \right] \nonumber\\
&&\beta {\cal F}' = \left({\beta \over 2}\right) \int d^3x~d^3x'~dn~dn'~
  \rho_t(\vec x,n) {V}(\vert \vec x - \vec x'\vert) \rho_t(\vec x',n')
  \nonumber\\
&&~\nonumber\\
~
\label{an0}
\end{eqnarray}

\begin{figure}
\includegraphics[width=4in]{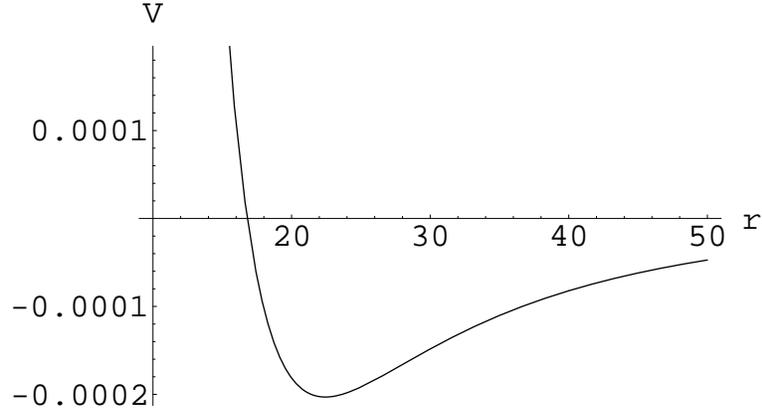}
\label{fig1}
\caption{This is a plot of the effective potential derived in the
  paper. Representative parameters were used: $\epsilon=80$,
$\lambda_e=3.4 \AA$, $\lambda = 10 \AA$.}
\end{figure}

One sees that the effective potential $V$, exemplified in
Fig.\ref{fig1} is given as follows:

\begin{equation}
{V}(\vert \vec x - \vec x'\vert) = \
{\exp\left(-\kappa_e \vert \vec x - \vec x'\vert\right) \over \vert \vec x - \vec x'\vert} -
{\exp\left(-\kappa \vert \vec x - \vec
      x'\vert\right) \over \epsilon~\vert \vec x - \vec x'\vert}
\label{an1}
\end{equation}

Observe that integrating over the solvent degrees of freedom has led
to an attractive (negative) supplement to the original short-ranged
screened Coulomb interaction.
The electrical polarization of the water, signified by $\epsilon$ serves to
weaken this attractive part of the potential. Given that the solvent
we consider is water, which has a natural $pH$ of 7, one always has an
associated non-zero screening length viz., $\lambda$.

The various parameters used in model, e.g. $q_+(n)$, $\lambda$,
$\lambda_e$ are inter-related.
This is because changing the ionic concentration in the solvent affects
not only $\lambda$, the solvent screening length, but also the amount
of adsorbed charge $q_+(n)$ and hence the screening length $\lambda_e$
along the chain as well.
And it may be prudent initially to appeal to experiment in order to
assess their magnitudes.

\begin{figure}
\includegraphics[width=4.0in]{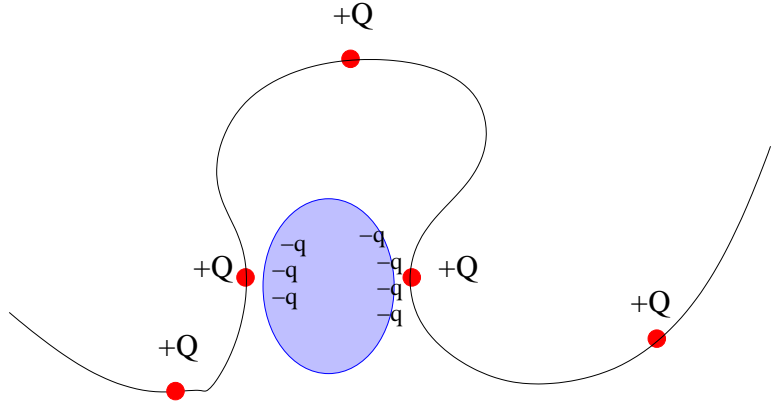}
\label{fig2}
\caption{A schematic view of the correlation effect which leads to an
  effective attraction between like charges on the chain.
Opposite charges in the solvent create a region which attracts like charges.}
\end{figure}

The physical notion behind the effective potential is analogous to what
happens in electron-phonon physics in metals, when the attraction
between electrons and phonons (motion of positive ions) leads to an
effective attraction between the electrons themselves.  This effect can
be depicted pictorially in Fig.\ref{fig2}.

The approach taken in this paper, viz., integrating over the solvent
degrees of freedom, is complementary to the approach
of Manning\cite{manning}, in which the focus rests on the
electrostatic potential of the solvent.  The polyelectrolyte,
represented by an infinite line charge, was shown by Manning to polarize the
surrounding ionic solvent.  There exists a thin annular cylinder
around the rod which attracts counter-ions, allowing for condensation
under certain conditions.  Manning's theory is a mean-field approach.
The approach developed in this paper is
more general. 
The present theory accounts for fluctuations around the mean-field approximation.
It allows for a general description of the physics of
polyelectrolytes, and is not restricted to a description of
condensation alone. 

Furthermore, as suggested earlier, if one takes the segment length $b$ 
to represent the correlation length of a semi-rigid
polymer, we expect the model to reproduce the overall phenomena
associated with semi-rigid polyelectrolytes.


\section{Regimes of polyelectrolyte behavior}

The effective potential derived in the previous section permits an
insight into the variety of behavior that polyelectrolytes in solution
can display. In order to do that, it is useful to add a constraint to the energy
functional described in the previous section, viz., ${\cal F}_{eff}$
to conserve the number of monomers on the polymer:

\begin{eqnarray}
\beta {\cal F}_{eff} &&\to \beta {\cal F}_{eff} + \beta \Delta {\cal
  F}_{\mu} \nonumber\\
\beta \Delta {\cal F}_{\mu} &&= -\mu \int dn \int d^3x \vert \psi(\vec
x,n)\vert^2 \nonumber\\
\mu &&= 2 \pi N {\ell}_b c_0 (\lambda_e^{2}-\lambda^{2}
\epsilon^{-1})\nonumber\\
{\ell}_b &&=\beta {(q_+-\vert q_-\vert)^2}
\label{uhat1}
\end{eqnarray}

where $\ell_B$ is the Bjerrum length, $c_0$ is the number density of
monomers, $N$ is the chain length, and the chemical potential is
obtained by minimizing the energy functional in Eqn.\ref{uhat1} in the
limit that the wave amplitude $\psi$ is independent of $\vec x,~n$.
It has also been assumed for simplicity that there is a uniform charge
distribution along the chain.

By convention, the chemical potential $\mu$ represents the energy
required to add a segment to the system. Hence, upon examining
Eqn.\ref{uhat1}, one sees three possible regimes.

{\em Melt:} When, $\mu > 0$, the polymer behaves as a melt, when
repulsion dominates, and it costs energy to add a segment to the
system.
This occurs when the screening length along the chain 
$\lambda_e > \lambda/\sqrt \epsilon$.
This is similar to Manning's discussion of polyelectrolytes.

{\em $\Theta$ point:} The monomers are perfectly miscible with each
other when the chemical potential $\mu=0$.  
This occurs when $\lambda_e = \lambda/\sqrt \epsilon$.

{\em Condensation:} When the screening length along the chain 
$\lambda_e < \lambda/\sqrt \epsilon$, the chemical potential becomes
negative, indicating that the system encourages the addition of
monomers. It is natural to identify this regime as that of
condensation.
Our criterion is similar to that of Manning\cite{manning}, 
but it requires a consideration of
fluctuations around the Debye-Huckel approximation.
In this sense our approach bears a greater resemblance to the analytic
arguments in the paper of Jensen et al\cite{jensen}.

As discussed earlier in the paper, if we assume that the native
charge on the chain is negative, viz., $q_- < 0$ and corresponds to the
charge of a single electron, then clearly, we require $q_+$ the charge
on the condensed counter-ion to be multi-valent, as otherwise the net
charge is null, leading to a neutral polymer, and a discussion of
condensation becomes moot. It is beyond the capability of the current
theory to address the difference in the condensation behavior of
{\em different} types of multi-valent molecules carrying the same
charge\cite{butler}.

Using SAX techniques, Butler et al\cite{butler} show that virus solutions
need a minimum diamine concentration for the onset of
condensation, as signaled by the emergence of a peak at a non-zero
wave-vector.  Moreover, they show that there also exists a maximum ion
concentration above which condensation ceases.
Equation \ref{uhat1} can be interpreted in light of these
experimental results. 
For low ion concentrations, $q_+$ the adsorbed counter-ion charge is
correspondingly small, and the screening length along the chain
$\lambda_e$ is very large. Since the dielectric constant of the
aqueous solvent is high ($\sim 80$), the chemical potential
$\mu = 2 \pi N {\ell}_b c_0 (\lambda_e^{2}-\epsilon^{-1}~\lambda^{2})$ can be greater
than zero, preventing condensation. 
As the ion concentration increases, the screening lengths along the
chain and  in the solvent decrease. When this change occurs in such a
way that $\lambda_e$ shrinks faster than $\lambda$, 
such that $\lambda > \epsilon \lambda$,
then the chemical
potential $\mu < 0$, leading to condensation.
Indeed this criterion is similar to the one derived on experimental
grounds by Butler et al.
For even higher ionic concentrations, $\lambda_e$ cannot shrink much
below the monomer length, and it is possible for the chemical
potential to become positive once again.  Condensation then ceases.
It may prove possible to adjust the parameters in this theory to
reproduce quantitatively the condensation data of Butler et
al\cite{butler}. Our aim here is to display that the theory is
simply capable of addressing experiments.
The cessation of condensation at relatively high ionic concentration
appears to be a new experimental result, and it is important to point
out that the theory in this paper can address this issue.

The theory presented in this paper is similar to that of Golestanian
and Liverpool\cite{golestanian}.
These authors use a phenomenological theory of semi-rigid chains
decorated with charges.  The only interaction in their theory is that
between these charges.  The presence of a solvent is implicitly
acknowledged through an effective screening length.  In our theory, it
would correspond to neglecting the terms ${\cal F}_{solvent}$ and
${\cal F}_{polymer-solvent}$.
They show that fluctuations in the shape of a semi-rigid rod lead to a buckling
instability.  
These authors too find cessation of condensation upon increasing the
salt concentration sufficiently, by 
considering the behavior of the correlation length.

The current theory can accommodate the zipper-like motif invoked by
Kornyshev and Leiken\cite{kornyshev}.
Our approach can be construed as providing an analytic background to
the numerical simulations of Stevens\cite{stevens} of
bead and spring polyelectrolytes in a solvent.

The onset of condensation in biopolymers, based on the assumption of
entirely rigid polyelectrolytes, has been shown by 
Kholodenko\cite{kholodenko} to be analogous to the melting of vortices
in Thouless' two-dimensional model 
of a glass. In the future, one could attempt to investigate aspect
further, using the current formalism, which does not assume rigid
rods, as follows.  Rather than integrating over the the electrostatic
potential of the solvent, one could attempt a functional integration
over the polymer degrees of freedom.  In this case, extremizing the
new effective functional with respect to the electrostatic potential
will lead to a generalization of the Kholodenko model to
non-rigid polyelectrolytes in solution.

\begin{figure}
\includegraphics[angle=0,scale=0.6] {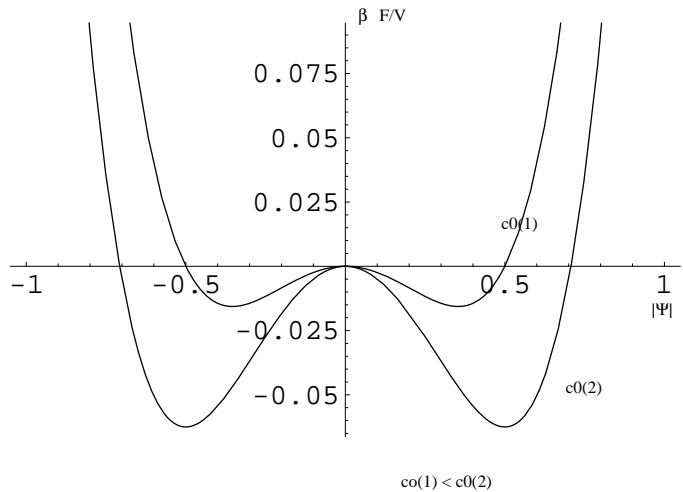}
\label{fig3}
\caption{A schematic of the double well potential given by the energy density $\beta f_{eff}$,
in arbitrary units.  Notice that the height of the potential barrier
decreases  as the concentration of monomers decreases, allowing for
more fluctuations between the minima.}
\end{figure}

$c_0 \to 0^+$:
In addition to the three regimes discussed above, there is one more important regime
within the melt phase which merits attention, and permits a connection
with the scaling arguments first derived by de Gennes\cite{deG1}.
Consider the energy density $\beta {f}_{eff}$ for a homogeneous
$\psi$:

\begin{equation}
\beta {f}_{eff}= (1/2) N \alpha \hat V(k=0) (\vert \Psi\vert^4 -c_0 \vert
\Psi\vert^2)
\label{2-well}
\end{equation}
where $\alpha = \ell_b/b$, 
where the dimensionless number concentration $c_0$ can be expressed in
terms of the actual monomer number concentration $\tilde c_0$ and the
Kuhn length $b$ as $c_0 = \tilde c_0 (b/\sqrt 6)^3~$
Thus when $\tilde c_0 (b/\sqrt 6)^3 <<~ 1$, one sees that the
double-well structure displayed in the next figure is just beginning
to form. 
This is the traditional signal for a second-order phase transition
(see Fig.\ref{fig3}).
Physically this occurs because the mean free path between collisions
is very large in the low number concentration limit, permitting large
fluctuations, the hall-mark of second order phase transitions.
Note that the criterion above encompasses the arguments given by Freed
\cite{freed}, where it is shown that phase transition-like behavior
occurs in the $b \to 0$ limit.

\section{Renormalization Group calculations in the melt regime}

We shall now show that the polymer model considered in
this paper exhibits a fixed point of the Renormalization Group
transformation,
reflecting the large fluctuations expected in the $c_0 b^3 \to 0^+$
(melt) regime.
Since a Renormalization Group
fixed point in the long wavelength limit (for long chain lengths) is
always independent of the details of the interaction, 
it is expected that our model of realistic
interactions should yield results identical to the excluded volume
model usually considered in polymer theory. 
Nevertheless, it is useful to derive this result explicitly so that it
displays the validity 
of the field theoretic approach employed in this paper.
Further, we shall compute a reasonably accurate value for the Flory
exponent, obtained when the chain length $N$ is very large.
In what follows, we shall use a natural system of units, in which the
unit of length will be taken to be $b/\sqrt 6$, where $b$ is the
segment length. As discussed earlier in the paper, $b$ is the monomer
length for a flexible chain. $b$ could be taken to be the correlation
length for a semi-rigid chain.

Finally, note that the model is analogous to the
theory of dynamical second order phase transitions\cite{hohenberg},
with the segment label $n$ playing the role of time.
The model has four independent variables,
viz., three spatial dimensions and an additional label for the
location of a segment along the chain.


Towards that end, let us consider the first order correction to the
bare vertex (the basic interaction displayed in Eqn \ref{interact0}).
It corresponds to the first order polarization diagram in many-body physics.
It represents the generalization of the correction that occurs in an excluded
volume interaction model\cite{shirish0}.

\begin{figure}
\includegraphics[scale=0.75]{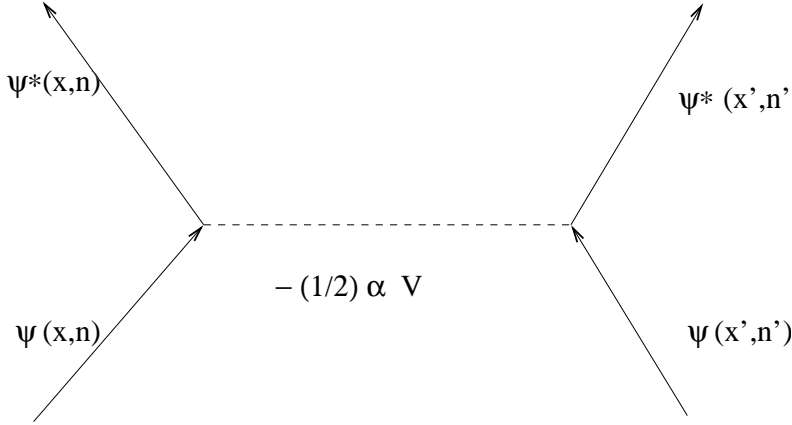}
\label{fig4}
\caption{
The Feynman diagram indicating schematically the bare vertex, whcih
represents the interction term in the energy functional.
It displays two segments interacting with each other at a distance.}
\end{figure}

The result depicted in this figure can be encapsulated as the renormalization of the coupling
constant $\alpha$ (using a length scale of $b/\sqrt 6$):

\begin{eqnarray}
\alpha \to \alpha_R(q,\omega) &&= \alpha~(1 + \alpha
\Pi_0(q,\omega) \hat V(q) )\nonumber\\
\Pi_0(q,\omega) &&= 
    -\int~ {d^3 q' \over (2 \pi)^3}~\int~{d \omega'\over 2 \pi}~\
 \hat G_0(q',\omega)~\hat G_0(\vert \vec q~'-\vec q
 \vert,\omega'-\omega)
\label{ar00}
\end{eqnarray}
where the Fourier transform $\hat G_0(p,\omega)$ of the unperturbed
propagator (Green's function)  defined in Eqn. \ref{one}, is given by:

\begin{equation}
\hat G_0(p,\omega) ={1 \over -i \omega + p^2}
\end{equation}

Now $\Pi_0(q,\omega)$ represents screening of the
interaction between monomers due to intervening segments.
Thus it is tacitly assumed that
we are in the long-wavelength regime, and the chain length is large,
as it is in this regime that one expects screening effects to
dominate.
Screening is clearly unimportant for extremely short chains, as the
probability of finding intervening segments in this case is
vanishingly small.

Upon performing the frequency integration in Eqn. \ref{ar00} by the method of residues,
by closing the requisite contour in the lower half-plane:

\begin{equation}
\Pi_0(q,\omega) =-~\left({2\over (2 \pi)^2}\right) {\cal R}e~\left[ \int dp p^2
\int_{-1}^{+1} dx~{1\over i \omega - 2 p q x} \right]
\label{ar1}
\end{equation}
where ${\cal R}e$ implies the real part of the quantity on its
right hand side is to be considered.
The polarization $\Pi_0(q,\omega)$ can be recast as:

\begin{equation}
\Pi_0(q,\omega) = -~\left({1\over 2 \pi^2 q}\right) \
\int_0^{\infty} d\lambda \lambda^{-1} \cos(\omega \lambda) \
\int dp p (\exp(-\zeta p) - \exp(-2 p q \lambda))
\label{ar2}
\end{equation}
where $\zeta \to 0^+$, and the factor of $\exp(-\zeta p)$ has been
inserted into Eqn. \ref{ar2} to achieve convergence.
We now invoke the integral representation of a gamma function, viz., 
$\int_0^{\infty} dx x^p \exp(-a x^m) = m^{-1} a^{-(p+1)/m}
\Gamma((p+1)/m)$ to obtain closed form expressions for the integrals
in Eqn. \ref{ar2}.
However this gives rise to divergent factors such as $\Gamma(-2)$.
These divergences can be {\em cured} via different methods, such as
Wilson's renormalization scheme.
The renormalization scheme to handle {\em ultra-violet} divergences
utilized in this paper follows more closely 
the method of dimensional
regularization\cite{ramond} of t'Hooft and Veltman.
This technique allows one to
separate out the infinities, by setting $p \to p - \epsilon$, $\epsilon
\to 0^+$, for $p+1/m$ equal to a negative integer, in the expression
above for the integral representation of the gamma function..
This allows us to express the polarization function as:

\begin{equation}
\Pi_0(q,\omega) = {-1\over (2 \pi)^2 q} \zeta^{-2}
+{\omega^{2-\epsilon}\over (4 \pi)^2 q^3} \Gamma(-2+\epsilon)
\label{rnn1}
\end{equation}

Using the Macclaurin series for $\Gamma(-2+\epsilon)$:
\begin{equation}
\Gamma(-2+\epsilon) \approx (1/2) \left( \epsilon^{-1} + \psi(3)  +
  {\cal O}(\epsilon)\right),
\end{equation}
one can isolate the divergent portions
of the polarization function as being proportional to $\zeta^{-2}$ and
$\epsilon^{-1}$.  These infinities can be canceled by the invocation
of appropriate counter-terms in ${\cal F}$, as is normally done in
field theory.  This allows us to retain only the finite portion of the
polarization function:

\begin{eqnarray}
\Pi_0(q,\omega) &&= -~{f(\omega)\over q^3}\nonumber\\
f(\omega) &&= -\left( \omega\over \sqrt 2 \pi \right)^2~\
\left( \ln(\vert \omega \vert) -3/2 + \gamma \right)
\label{ar3}
\end{eqnarray}
where $\gamma$ is Euler's constant, numerically close to $0.5772$.

A test of the renormalization scheme used to handle ultra-violet
divergences here is to calculate an
observable or a known quantity such as the Flory exponent.
This will be done shortly, after a thermodynamic fixed point has been
identified. This fixed point is associated with the second order phase
transition discussed in the previous.

One must go to the long wavelength limit in order to invoke the
Renormalization Group.
This Renormalization Group applies to the infra-red limit ($q \to 0$),
and must be distinguished from the preceding discussion regarding the
behavior in the $q \to \infty$ limit.

It is convenient to redefine a scaled
coupling constant in the $q\to 0$ limit, with an associated
$\beta$-function
which yields the Renormalization Group flow:

\begin{eqnarray}
\hat \alpha_R(q,\omega) &&= \alpha_R ~\left({f(\omega) \hat V(q=0)\over
    q^3}\right) \nonumber\\
\beta(\hat \alpha_R) \equiv {\partial \alpha_R \over \partial \ln L}
  && = 3 (\hat \alpha_R - \hat \alpha_R^2)
\label{ar4}
\end{eqnarray}
where $q \equiv L^{-1}$.
In analogy with the theory of second order phase transitions,
scale invariance of the coupling constant must be invoked for a proper
description of the physics.
It follows from Eqn. \ref{ar4} that
$\beta(\hat \alpha_R)$ has a non-trivial fixed point at
$\hat \alpha_R \equiv \hat \alpha_R^*$,
or at $\alpha_R \equiv \alpha_R^*$:

\begin{eqnarray}
\hat \alpha_R^* &&= 1 \nonumber\\
\alpha_R^* &&= {q^3 \over f(\omega) \hat V(q=0)}
\label{ar5}
\end{eqnarray}

It follows from Eqn.\ref{ar5} that in the long wavelength limit, in
the $c_0 \to 0^+$ regime, the interaction has the following universal
form, independent of the particular interaction potential we started
off with:

\begin{equation}
\alpha \hat V(q\to 0) \to -~\frac{2 \pi^2 q^3}
    {\omega^2 \left( \ln(\vert \omega \vert) -3/2 + \gamma \right)}
\label{ar6}
\end{equation}

The same result could have been obtained in a more mundane fashion by
using the so-called Random 
Phase Approximation (RPA) in which one traditionally sums up bubble
diagrams of all orders. Nonetheless, the Renormalization Group
approach is essential in that it
explicitly manifests the scale invariant properties of the system.

Scaling arguments will be utilized to obtain a value of the Flory
exponent.
These are provided in lieu of the conventional Renormalization Group
argument based on scaling transformations.  That treatment is
based on the scaling properties of the simpler $\phi^4$
Ginzburg-Landau functional. In the present paper, we have a more
realistic interaction potential viz., $V$ to deal with.
Consequently, the straighforward scaling properties of the $\phi^4$
excluded volume theory are no longer valid.  A more complex set of
scalings, involving the screening lengths are required to maintain the
scale invariance of the current theory. But this implies a restriction
on the class of polyelectrolytic solutions that can be considered,
conforming to the scaling transformations.
The alternate derivation offered below suffices to yield the Flory
exponent fairly accurately.

Consider scaling behavior of the energy functional for a Gaussian chain:

\begin{equation}
\beta {\cal F}_0 = \int {d \omega \over 2 \pi} ~\int {d^3k\over (2 \pi)^3}\
\hat \psi^*(k,\omega)(-i \omega + k^2) \hat \psi(k,\omega)
\label{rn1}
\end{equation}

The diffusion-like, self-similar form permits Functional to be scale invariant under
the following transformations:

\begin{eqnarray}
\omega &&\to \omega/{\cal N} \nonumber\\
k &&\to k/\xi({\cal N}) \nonumber\\
\xi({\cal N}) &&= {\cal N}^{\nu_0} \nonumber\\
\hat \psi &&\to \hat \psi \xi^{5/2}({\cal N})
\label{scl1}
\end{eqnarray}
where $\nu_0 = 1/2 $.
The relation between $\xi$ and ${\cal N}$ is identical to the relation
between the radius of gyration $R_g$ and $N$. Hence we shall identify
the length scale $\xi({\cal N})$ with the radius of gyration.

Let us now consider the case when the interaction viz., $V$ is turned
on between the segments.
The
behavior of the energy functional in the long wavelength limit ($q\to 0, \omega \to
0$) is dominated by the behavior of the effective interaction at the
Fixed point of the Renormalization Group transformations (see Eqn.\ref{ar6}):

\begin{equation}
{\cal F}^*\sim \int {d \omega \over 2 \pi} ~\int {d^3k\over (2
  \pi)^3} ~\hat c^*(k,\omega)~\left({k^3 \over \omega^{2- \pi^{-2}}}\right) ~\hat c(k,\omega)
\label{scl2}
\end{equation}
where we have used $x^\epsilon \approx 1 + \epsilon \ln x$, $\epsilon
<<1$ to convert a logarithm to a power law.

This functional exhibits invariance under:

\begin{eqnarray}
\omega &&\to \omega/{\cal N} \nonumber\\
k &&\to k/\xi({\cal N}) \nonumber\\
\xi({\cal N}) &&= {\cal N}^{\nu} \nonumber\\
\hat \psi &&\to \hat \psi \xi^{3/2}({\cal N})
\label{scl3}
\end{eqnarray}
where
$\nu = 2/3-1/(3 \pi^2)\approx  0.63$.

This estimate of the radius of gyration holds for asymptotically large
chain lengths $N$, when the concentration of
monomers $\tilde c_0 << (\sqrt 6/b)^3$.

If one accepts the value of $\nu \approx 0.5889$ as obtained within
the framework of the Edwards' model, then the result in this paper is
accurate to about $92\%$. 
On the other hand, H. Kleinert\cite{kleinert} has shown that when the asymptotic
series in the $\epsilon$ expansion is handled appropriately, the value of
$0.5889$ is replaced by $3/5$.
Kleinert then uses field theoretic methods and the replica trick to
rederive a value of $\nu \approx 0.62$.
The value we obtained, $\nu \approx 0.63$ is fairly close to the
universal value for $\nu\approx 0.634$ accepted in phase transition
theory.

This value is quite different from the one derived by Flory\cite{flory-el} and
Katchalsky\cite{kat} for the case of an unscreened Coulomb interaction, which
yielded a linear dependence on the chain length.


\section{Radius of gyration for short chains}

\begin{figure}
\includegraphics[width=4.5in]{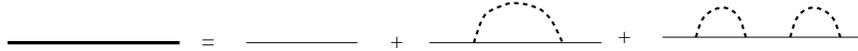}
\caption{A diagrammatic depiction of the approximation used in the
  calculation. The thick line indicates $\hat G$. The first term on
  the right hand side of the equation refers to $\hat G_0$. The second
  term refers to the {\em exchange}-like diagram referred to in the
  text. The final term is the second order correction.}
\label{fig5}
\end{figure}

For the case of short chains, when the coupling constant 
$\alpha = \sqrt 6{\ell_b}/b < 1$, 
one can compute the lowest order Feynman diagram to estimate the
self-energy, which renormalizes the Green's function $\hat G_0(k,\omega)$. 
This estimate is correct when $\alpha < 1$ and when the chains length
is relatively short.  This is because screening is expected on
physical grounds to be unimportant for short chains.
It turns out that there are two terms of ${\cal O}(\alpha)$ in a
perturbative expansion of the self-energy.  The first one, the
tadpole diagram can be renormalized away in the usual fashion by the
addition of a counter-term.  The remaining term is reminiscent of the
exchange diagram in many-electron physics (as shown in Figure\ref{fig5}).

Using the Feynman-Dyson technique, repetitions of this diagram can be summed as a
geometric series to infinite order.
This exchange diagram contribution to the self-energy can be expressed
in closed form as follows:

\begin{equation}
\Sigma_{exch}(k,\omega) = -\alpha \int \frac{d^3 q}{(2 \pi)^3} \hat
G_0(k,\omega) \hat V(\vert \vec k - \vec q \vert)
\label{exch}
\end{equation}
where it is tacitly assumed that we are in the low concentration limit.
Since our focus is on estimating the radius of gyration for relatively
short chains, we need only to evaluate the self-energy in the
long-wavelength limit, viz., $k \to 0$. To facilitate this, the
angular integrals in Eqn.\ref{exch} can be done and the contributions
to ${\cal O}(k^2)$ of the integrand can be written explicitly:

\begin{eqnarray}
&&\int_0^{2 \pi}~ d\phi~\int_0^{\pi}~d\theta~ V(\sqrt{k^2+q^2 -2 k q
  \cos \theta}) \approx \nonumber\\
&&8\,{\pi }^2\,\left[
     \frac{1}{q^2 + {{{\kappa }_e}}^2} -\left( \frac{1}
        {\epsilon \,\left( q^2 + {\kappa }^2 \right) } \right)
                \right]  + \nonumber\\
&&\left(\frac{8\,{\pi }^2\,k^2}{3}\right)~\left[ \frac{3}
        {\epsilon \,{\left( q^2 + {\kappa }^2 \right) }^2} - 
       \frac{3}{{\left( q^2 + {{{\kappa }_e}}^2 \right) }^2} + 
       q^2\,\left(
          \frac{4}{{\left( q^2 + {{{\kappa }_e}}^2 \right) }^3}-
   \frac{4}{\epsilon \,{\left( q^2 + {\kappa }^2 \right) }^3}
          \right)  \right]
\label{ang}
\end{eqnarray}

The subsequent $k-$integration can be performed with the aid of
Mathematica.  The results are long and cumbersome and not much is
gained by stating the expressions explicitly, other than to say that
the integrals contain logarithmic terms of the form $\ln \omega$.
Formally, the next step is to evaluate the correlation function:

\begin{equation}
{\cal G}(k,N) \approx \int \frac{d \omega}{2 \pi} ~\frac{\exp(-i \omega
  N)}{-i \omega + k^2 - \Sigma_{exch}(k,\omega)}
\label{exch1}
\end{equation}

This integral could be
evaluated using the method of residues, by closing the contour in the
lower half plane if the root(s) of the denominator
could be located.
Since our interest is in the $k \to 0$ limit, it is possible to
estimate the root of the denominator perturbatively:

\begin{equation}
\omega_{root} \approx = - i k^2 + i \Sigma_{exch}(k,\omega = -i k^2) +
{\cal O}(\alpha^2)
\label{root}
\end{equation}

Before the residue can be calculated, the presence of the logarithmic
terms referred to above imply the existence of a branch cut in the
complex $\omega$ plane from $-\infty$ to $0$ along the real axis.
Hence the contour has to distorted slightly into the lower half-plane
along the negative real 
axis to avoid this branch cut, so that the only singularity enclosed
by the contour in the lower half-plane is at the root given by
Eqn.\ref{root}.
It then follows, using Mathematica, that in the long wavelength limit:

\begin{eqnarray}
{\cal G}(k,N)/{\cal G}(k=0,N) &&\approx \exp (-i k g(N) )\nonumber\\
g(N) &&=\alpha~(\lambda_e^4 - \epsilon^{-1}~\lambda^4) b^{-4}
\label{sh1}
\end{eqnarray}
where $\lambda_e$ and $\lambda$ are the screening lengths along the
chain and in the solvent, respectively.

The experimentally measurable structure factor can then be evaluated:

\begin{eqnarray}
\hat S(k) &&\approx N ( 1- (1/3) k^2 R_g^2(N) + {\cal
  O}(k^4))\nonumber\\
R_g &&= (b_{effective}/\sqrt 6)~ N \nonumber\\
b_{effective}&&= (18 \alpha)~
\left[\left(\frac{\lambda_e}{b}\right)^4 - \epsilon^{-1} \left(\frac{\lambda}{b}\right)^4\right]~b
\label{sh2}
\end{eqnarray}

\begin{figure}
\includegraphics[scale=0.75]{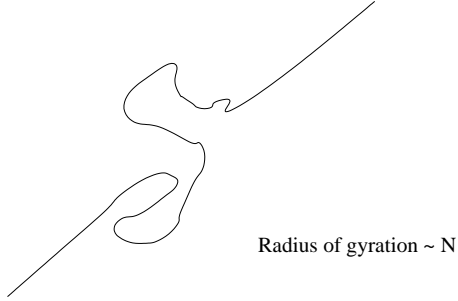}
\label{fig6}
\caption{ Schematic depiction of the linear behavior of the radius of
  gyration for short chains.  Note that the Kuhn length can be shorter
  than the segment lenght due to a fluctuation effect.}
\end{figure}

Note that $R_g(N)\sim N$ in this regime, as displayed in
Fig.\ref{fig6}.   Moreover, it turns out that
$R_g(N)$ is extremely sensitive to the parameters in the potential.
This is in contrast to the asymptotic regime encountered in the
previous section, where universal behavior was found.
X-ray or neutron scattering experiments could be performed to test the
predictions presented in this section.

One could interpret $b_{effective}$ in Eqn.\ref{sh2} as an effective correlation
length and suggest $b_{effective} \to 0$ as a signal for buckling,
and hence for condensation.  This interpretation is equivalent to the
chemical potential considerations presented earlier in the paper, and
is similar to the approach of Golestanian and Liverpool\cite{golestanian}.


\section{Finite concentration of segments}

The discussion in the preceding two sections has been primarily
focused on behavior in the melt regime, when the number concentration
of segments is vanishingly small.  In this case, fluctuations were
accounted for, and the radius of gyration for both long and short
chains was computed.  As the concentration of segments increases, the
segments become packed increasingly closer, thereby
decreasing fluctuations in the system.  Mean field approximations,
obtained by extremizing the energy functional can
then be invoked to obtain insight into the physics.

In a previous paper,\cite{shirish0} which utilized an excluded volume interaction
model, we showed how tube-like structures can be obtained.
If one restricts attention to obtaining an envelope of structures
obtained in the various regimes delineated in section III, then a
similar technique provides useful insight in the current approach as
well. The basic idea, designed to ease computation, is to replace the
short-ranged potential by an effective delta-function
pseudo-potential.  The effective coupling constant which characterizes
the pseudo-potential can be positive or negative, depending on the
value of the chemical potential $\mu$.  As discussed earlier, the
value of the chemical potential is an average way of determining
whether the attractive part or the repulsive part dominates the
behavior of the system.  

The advantage of this method is that it
yields the correct average behavior of
the system for a reasonably small effort.  
The disadvantage is that if one is interested in details
of the structures which change on the spatial scale less than the one
over which the
interaction potential varies, then one must resort to vastly more
detailed calculations.
The approximation consists of the following replacement:

\begin{eqnarray}
&&(1/2) \alpha \int \int dn d^3x~dn' d^3x'~ |\psi(\vec x,n)|^2 V(\vec x
- \vec x') |\psi(\vec x',n')|^2 \to \nonumber\\
&&(1/2) \alpha \hat V(k=0)~\int dn
d^3x~ \int dn'~|\psi(\vec x,n)|^2~|\psi(\vec x,n')|^2
\label{local1}
\end{eqnarray}

Extremization of the functional leads to:

\begin{equation}
({\partial \over \partial n} -\nabla^2 +  \alpha \hat V(k=0) |\psi(x,n')|^2 - \mu ) \psi(x,n)
= 0
\label{local2}
\end{equation}

For cases when the segment label $n$ is not physically relevant, this
equation reduces to:

\begin{equation}
(-\nabla^2 \pm~( |\psi(x)|^2 - 1) ) \psi(x)=0
\label{local3}
\end{equation}

where the distances are scaled in units of $(b/\sqrt 6) (\alpha |\hat
U_{Morse}(k=0)|)^{1/2}$, and the amplitude $\psi$ has been scaled by
$c_0^{-1/3}$ so as to be dimensionless.
The positive sign in the non-linear partial differential equation
holds when $\mu > 0$ in the melt phase, and the negative sign when in the
condensed state ($\mu < 0$).

Associating $\mu < 0$ with a condensed state implies the existence of
coherent structures.  The first structure we investigate is one in
spherical geometry, with a larger amplitude at the center than towards
the edges.  The ordinary differential equation that requires a
solution is:

\begin{equation}
-{1\over r^2}~{\partial \over \partial r}\left(r^2 {\partial \psi(r) \over
    \partial r}\right) - (|\psi(r)|^2 -1) \psi(r)=0
\label{sph1}
\end{equation}

It turns convenient for numerical purposes to use $y=1/r$, so that
Eqn.\ref{sph1} becomes:

\begin{equation}
-y^4 \Psi''(y) -(\Psi(y)^2 -1) \Psi(y)=0
\label{sph2}
\end{equation}

\begin{figure}
\includegraphics[width=4in]{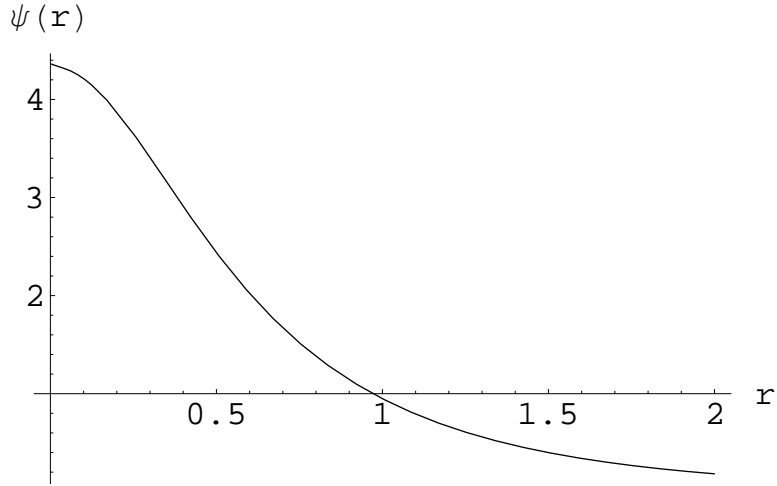}
\label{fig7}
\caption{ Displayed is the radial profile of a spherical coherent
  structure obtained for $\mu < 0$.  The co-ordinates are
  dimensionless.}
\end{figure}

For $y \to 0$, which is the same as $r \to \infty$, the wave amplitude
is expected to decay, so that the non-linear term vanishes, and
yields $\Psi(y) = a y \exp(1/y)$.  One can now integrate numerically
from some $y=y_{minimum}$ to $y=y_{maximum}\equiv r_{minimum}$, adjusting $a$ such that the slope
of the amplitude vanishes at $y_{maximum}$.
The result is shown in Figure \ref{fig7}.

Another interesting structure that we have investigated is a toroidal
structure.  The interest in this structure arises from the fact that
strands of DNA in an ionic solution condense into such shapes under
appropriate solvent conditions\cite{golan,veena}.  One may construe the model above
(Eqn.\ref{local3}) as an effective model for DNA in solution.
The co-ordinate surfaces in toroidal geometry are\cite{margenau}: (i) planes through
the z-axis, represented by an azimuthal angle, (ii) spheres of varying
radii centered up and down the z-axis, and finally, (iii) {\em tores},
or anchor rings around the z-axis, labeled by the location of their
centers at a distance $a$, and cross-sectional and axial radii $a~{\rm
  csch}
\eta$ and $a~\coth \eta$ respectively, for $\eta=$ constant, $\eta$ being the
toroidal co-ordinate. Since we seek solutions with toroidal symmetry, the only
independent variable we need to consider is $\eta$, which leads to a
non-linear ordinary differential equation:

\begin{equation}
2 \sinh(\eta/2)^2 (\tanh(\eta/2) \psi'(\eta) - 2 \sinh(\eta/2)^2
\psi''(\eta)) - a^2 (\psi(\eta)^2 -1) \psi(\eta) = 0
\label{tor1}
\end{equation}

The radius $a$ of the torus is determined self-consistently as the
value which yields a zero slope but non-zero amplitude at $\eta \to
\infty$.  This is the center of the {\em doughnut}.
Again, since we are looking for a doughnut shaped object, it follows
that the slope of the function should be zero at a value of $\eta$ at
which the cross-sectional radius of the tore is $a~{\rm csch} \eta = a$,
i.e., when $\eta = {\rm arcsinh} 1 \approx 0.8813$.
At this point
the the z-axis is a tangent to the tore at the origin, and physical
consideration implies a zero slope for continuity.
The value of the wave amplitude is taken to be zero at $\eta={\rm
  arcsinh} 1$, i.e., the origin.
A shooting method was employed where
$a$ was varied 
iteratively until a solution with a zero slope at $\eta \to \infty$
was obtained numerically.  Operationally, the equation was integrated
to some large value of $\eta$.
The result is displayed as contours in Figure VIII in the $x-z$ plane..
In effect a doughnut shaped structure is obtained, whose hole is
partially filled. The energy of this structure is identical to that of the
spherical blob displayed in Fig. \ref{fig7}.

\begin{figure}
\includegraphics[width=4in]{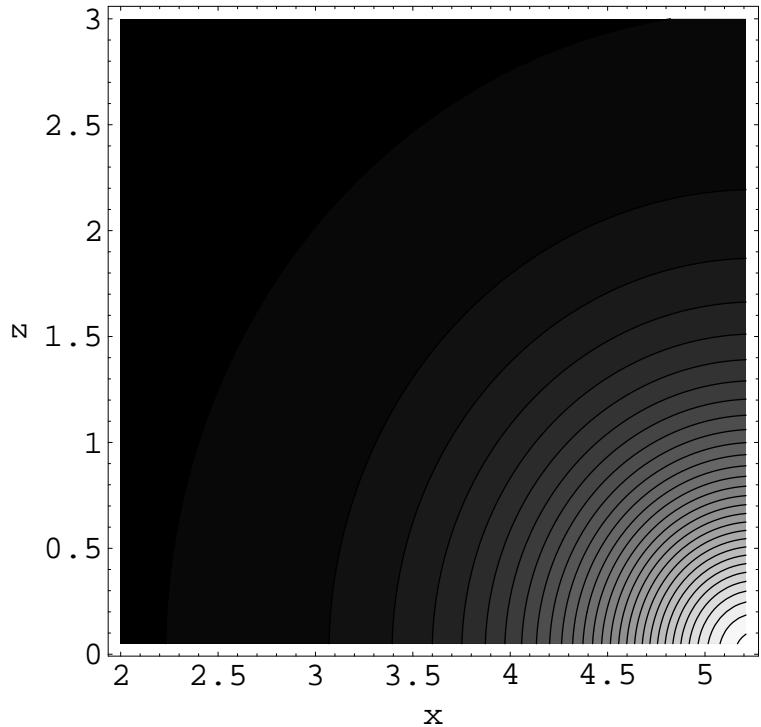}
\label{fig8}
\caption{ Displayed are the contours of a toroidal structure for $\mu < 0$.
Note that the doughnut has a profile which indicates a changing
density with distance from the center. Note that the width of the
structure is approximately unity in dimensionless units.}
\end{figure}

These are two examples of coherent structures that are possible for
$\mu < 0$. They are both equally energetically favorable. 
As such, the approximations employed during these calculations are
applicable for large chain lengths, when it is known exoperimentally
that either spheroidal or toroidal configurtions are equally likely to
occur\cite{golan,veena}.
These examples do not constitute an exhaustive list.

Finally, we note from the previous figure, displayed in dimensionless
variables, that the width of the toroidal configuration is ${\cal
  O}(1)$, in the length scale used.  Based on this observation, one
can calculate the thickness of the toroid for various parameters, and
an example is given in Fig. IX. Experiments of the sort performed by Golan
et al\cite{golan} and by Butler et al\cite{butler} should be able to
provide experimental verification of our predictions.

\begin{figure}
\includegraphics[width=6in]{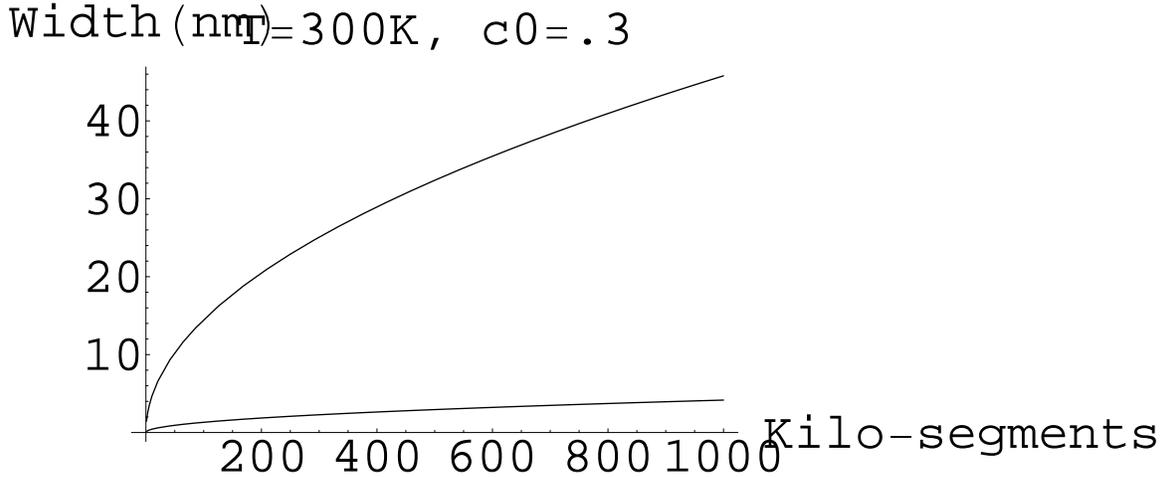}
\label{fig9}
\caption{ The width of the toroid is plotted in nano-meters
  for two different cases, as the chain length is varied, for
  two different screening lengths in the solvent. 
The upper curve coresponds to a screening lenght of $50 b/sqrt 6$.
The lower curve coresponds to a screening lenght of $10 b/sqrt 6$.} 
\end{figure}

\section{Conclusion}
The theory described in this paper develops a functional integral
technique to treat realistic interactions between segments of a polymer in a
realistic way, through the use of a finite-ranged repulsive-attractive
interaction potential.
Examination of the chemical potential led to a classification of
homopolymeric systems.  It was pointed out that such a classification
would be impossible with an a priori excluded volume interaction model.
Renormalization Group techniques were used to show that standard
concepts in polymer physics are recovered in the limit of low monomer
number concentration, for asymptotically long chains, in the melt state.
The radius of 
gyration for extremely short chains is also calculated, and is linear
in the chain length,
reminiscent of a semi-flexible chain.
When the chemical potential is negative, condensed structures are
shown to exist, both in spherical as well as toroidal geometry.
The predictions that follow from the theory presented in this paper,
viz., the short chain radius of gyration, the widths of the toroidal
configurations as functions of various experimentally accessible
parameters could be verified experimentally.

\section{Acknowledgments}

I would like to acknowledge useful discussions with Hank Ashbaugh, Tommy Sewell and
Kim Rasmussen, and thank Jack Douglas and Enrique Batista for
providing useful references.

This work was supported by the Department of Energy Contract no.
W-7405-ENG-36.


\end{document}